\documentclass[a4paper,12pt]{article}

\usepackage{cite}
\usepackage{pstricks}
\usepackage{multirow}
\usepackage{latexsym}
\usepackage{amssymb}
\usepackage{bm}
\usepackage{epstopdf}

\usepackage{graphicx}

\newcommand{\be}{\begin{equation}}
\newcommand{\ee}{\end{equation}}

\newcommand{\beq}[1] {\begin{equation}\label{#1} }
\newcommand{\eeq} {\end{equation} }
\newcommand{\bea}[1]{\begin{eqnarray}\label{#1} }
\newcommand{\eea}{\end{eqnarray}}

\def\beqn{\begin{eqnarray}}
\def\eeqn{\end{eqnarray}}

\def\beq{\begin{equation}}
\def\eeq{\end{equation}}
\def\bea{\begin{equation}}
\def\eea{\end{equation}}

\def\vs{\vspace}

\begin{document}
\vspace*{-0.2in}
\begin{flushright}
OSU-HEP-11-03\\
KANAZAWA-11-05\\
March, 2011\\
\end{flushright}

\vs{0.5cm}

\begin{center}
{\Large\bf Variations on the Supersymmetric $Q_6$ Model of Flavor}\\
\end{center}

\vspace{0.5cm}
\begin{center}
{\Large
{}~K.S. Babu$^{a,}$\footnote{E-mail: babu@okstate.edu}{},
{}~Kenji Kawashima$^{b,}$\footnote{E-mail: kenj@hep.s.kanazawa-u.ac.jp
} and
{}~Jisuke Kubo$^{b,}$\footnote{E-mail: jik@hep.s.kanazawa-u.ac.jp
}
}
\vspace{0.5cm}

{\em $^a$Department of Physics, Oklahoma State University, Stillwater, OK 74078, USA }
\vspace*{0.1in}

{\em $^b$Institute for Theoretical Physics, Kanazawa University, Kanazawa 920-1192, Japan
}

\end{center}

\begin{abstract}

We observe that a recently proposed supersymmetric model with
$Q_6$ flavor symmetry admits a new CP violating ground state.
A new sum rule for the quark mixing parameters emerges,
which is found to be consistent with data. Simple extensions
of the model to the neutrino sector suggest an inverted
hierarchical mass spectrum with nearly maximal CP violation
($|\delta_{\rm MNS}| \simeq \pi/2)$. Besides reducing the number
of parameters in the fermion sector, these models also provide
solutions to the SUSY flavor problem and the SUSY CP problem.
We construct a renormalizable scalar potential that leads to
the spontaneous breaking of CP symmetry and the family symmetry.
\end{abstract}

\newpage

\section{Introduction}

Non--Abelian discrete symmetries have found applications in explaining aspects
of the flavor question not addressed by the standard model (SM) of particle physics.
Restrictions imposed by such symmetries can lead to predictions for the
Cabibbo--Kobayashi--Maskawa (CKM) quark mixing angles in terms of the quark
mass ratios \cite{pakvasa}.  Such symmetries have been employed successfully
to generate a geometric structure for the leptonic mixing angles, independent
of the lepton mass ratios \cite{ma}.

The non--Abelian discrete symmetric structure $G$ appears
to bode well within the supersymmetric (SUSY) standard model, since the same symmetry
can also provide a natural solution to the excessive flavor change that occurs
with generic soft SUSY breaking terms  \cite{FCNC}.  The three families of quarks and leptons  will transform
as doublets or triplets of the group $G$, which would result in the degeneracy of
their masses.  Degenerate squarks (and sleptons) would alleviate the SUSY flavor
violation problem.  If the non--Abelian symmetry is continuous \cite{dine} and gauged \cite{bm}, there
is flavor violation arising from the $D$--terms of the flavor group \cite{murayama}.  Continuous
global symmetries are susceptible to explicit violation from quantum gravity.  Non--Abelian
discrete symmetries which have a gauge origin are free from these problems and deserve
special considerations \cite{seiberg}.

In Ref. \cite{bk} we presented a SUSY model based on the non--Abelian discrete group
$Q_6$ -- the binary dihedral group of order 12.  This group has two inequivalent doublet
representations, one real doublet, and one pseudo--real doublet, which can be handy for
model building.  In the flavor sector this symmetry
results in one prediction for for a combination of the CKM mixing parameters, which was shown to be consistent
with data.  The quark and lepton superfields are assigned to doublets and singlets of $Q_6$,
with the singlets identified as belonging to the third family.  In the $Q_6$ symmetric limit
the squarks of the first two families would be degenerate, which is sufficient to solve the
SUSY flavor problem.  Furthermore, by assuming that CP violation has a spontaneous origin,
this model also solves the SUSY CP problem.  Excessive CP violating processes arising from the
SUSY breaking sector are absent, since the parameters are all real.  Yet the model admits
CP violation in the quark mixing matrix.

One major purpose of the present paper is to show that the $Q_6$ model
studied in Ref. \cite{bk} admits a new minimum which violates CP, but leaves a new interchange
symmetry in tact.  By virtue of this interchange symmetry, we derive a new sum rule among the
quark mixing parameters and CP violation, which is found to be consistent with observations.
Such an interchange symmetry was present in Ref. \cite{bk} as well, but the new one presented here is
different, although it arises from the same Higgs potential.  We extend this symmetry to the lepton
sector and obtain interesting correlations between the neutrino oscillation parameters.
We also compare the predictions of the new minimum with those of the old, and in the
process update our old predictions.  We use the most recent values of light quark masses where the
errors have decreased significantly as a result of improved lattice calculations. We compare
the model predictions to the best fit values obtained in the SM as well as by including certain
new physics contributions in $B_{d,s}-\overline{B}_{d,s}$ mixings as obtained by the  CKFfitter group \cite{Lenz:2010gu}.
These new contributions
are motivated by certain discrepancies obtained in the SM CKM fits -- such as the differences of order 20\%
in the CP violation parameter $\overline{\eta}$ obtained from fits to $\epsilon_K$ and
$B \rightarrow J/\Psi K_S$ decay.  Small new physics contributions naturally arise in our
$Q_6$ based model.  For example, there are contributions to meson--antimeson mixing via SUSY box diagrams,
which may be important for the $B_{d,s}$ meson system
since the third family squark is not degenerate with the first two family squarks.

We also present a complete Higgs potential that leads to the spontaneous breaking of CP symmetry and the $Q_6$
flavor symmetry without leading to pseudo--Nambu--Goldstone bosons.  In addition to the
$Q_6$, a flavor universal $Z_4$ symmetry is introduced.
Owing to this $Z_4$, even after spontaneous symmetry breaking, an unbroken interchange symmetry
survives in the Higgs potential, for which there are two possible choices, denoted as ${\cal P}_{I,II}$.
These symmetries, along with $Q_6$, reduce
significantly the number of parameters in the fermion mass matrices. This reduction of parameters leads to a sum rule involving quark masses and mixings \cite{bk}. Moreover, CP violation has a spontaneous origin, which is perhaps more satisfying
than the usual assumption of explicit CP violation.  Nevertheless, the dominant source of CP violation in the quark sector is the Kobayashi-Maskawa mechanism.  The sum rule involving quark masses and mixings that has been  derived
relies on the spontaneous violation of CP.  With this, the problem of excessive
CP violation that generically exists in the soft SUSY breaking sector
can be solved in a rather simple way. Various phenomenological aspects of this model in minimum ${\cal P}_I$ have
been studied in Ref. \cite{bk,Kajiyama:2005rk,Babu:2009nn}.

The plan of the paper is as follows.  
In Sec. 2 we present the supersymmetric $Q_6$ model.  In Sec. 3 we analyze the predictions of model ${\cal P}_I$.
In Sec. 4 we provide the new model ${\cal P}_{II}$ and analyze its predictions for the quark mixing angles
and CP violation.  Sec. 5 discusses a simple extension of model ${\cal P}_{II}$ to the neutrino and charged lepton
sector and the resulting predictions.  In Sec. 6 we have our concluding remarks.

\section{CP invariant SUSY $Q_6$ model}

\subsection{$Q_6$ group theory and the Yukawa sector of the model}

We work within the context of supersymmetric standard model, with a non--Abelian flavor
symmetry $Q_6$ acting on the three families of quarks, leptons and their superpartners.
The group theory of $Q_6$ is discussed in detail in Ref. \cite{bk}.
We briefly recall its salient features
relevant for model building.  $Q_6$ is a binary dihedral group,
a subgroup of $SU(2)$, of order 12.  It has the presentation
\begin{equation}
\{A, B; A^6=E, B^2=A^3, B^{-1}AB=A^{-1}\}~.
\end{equation}
The irreducible representations of
$Q_6$ fall into $2, 2^{'}, 1, 1^{'}, 1^{''}, 1^{'''}$,
where the 2 is complex--valued but pseudoreal, while the $2^{'}$ is real valued.
The $\{1,~1',~1'',~1'''\}$ singlets form a $Z_4$ subgroup with
the $1$ and $1^{'}$ being real and the $1^{''}$ and $1^{'''}$ being
complex conjugates of each other. The group multiplication rules are given as
\begin{equation} 1'\times 1'=1,\ 1''\times 1''=1',\ 1'''\times 1'''=1',\ 1''\times 1'''=1,\
1'\times1'''=1'',\ 1'\times1''=1''' \end{equation}
\begin{equation} 2\times1'=2,\ 2\times1''=2',\ 2\times1'''=2',\ 2'\times1'=2',\ 2'\times1''=2,\ 2'\times1'''=2
\end{equation}
\begin{equation} 2\times2=1+1'+2',\ 2'\times2'=1+1'+2',\ 2\times2'=1''+1'''+2 \end{equation}
The Clebsch--Gordon coefficients for these multiplication can be found in Ref. \cite{bk}.

In Table 1 we list the $Q_{6}$ assignment of the quark,
lepton and Higgs chiral supermultiplets in our model,\footnote{Essentially the same model can be realized with any $Q_{2N}$
if $N$ is odd and a multiple  of $3$.} where
$Q, Q_3, L, L_3$ stand for the $SU(2)_L$ quark and lepton fields,  and $ H^u, H_3^u, H^d, H_3^d$
are the Higgs doublets. The $SU(2)_L$ singlet
supermultiplets for quarks, charged leptons and neutrinos are denoted by
$u^c, u^c_3,d^c, d^c_3, e^c, e^c_3$ and $\nu^c, \nu^c_3$.
Three pairs of Higgs doublets are introduced in order to generate
fermion masses directly in the presence of $Q_6$ symmetry using renromalizable couplings.
The singlet field $T_3$ is needed to generate the Majorana mass for $\nu_3^c$.
The other singlet scalar fields are needed to achieve spontaneous breaking of $Q_6$ symmetry
as well as CP symmetry without giving rise to pseudo--Nambu--Goldstone bosons.  This point
will be clarified in the next subsection.  Table 1 also shows a flavor universal $Z_4$ symmetry,
the purpose of which is to realize an unbroken interchange symmetry in the scalar sector even after
spontaneous symmetry breaking.  Such an interchange symmetry, for which we have two solutions,
enables us to predict one quark mixing parameter.

\begin{table}[h]
\caption{Particle content of the $Q_6$ model along with their transformation under
$Q_6 \times Z_4$.}
\vspace*{-0.05in}
$$\begin{array}{|c||c|c|c|c|c|c|c|c|c|c|c|c|c|c|c|c|c|c|c|c|c|}
\hline
~& \{Q,L\}  & \{Q_3,L_3\} & \{u^c,d^c,\nu^c,e^c\} & \{u_3^c,d_3^c,\nu_3^c,e_3^c\} &H^{u,d} & H_3^{u,d} & S & S_3 & T &T_3 &U\\
\hline
Q_6 & 2 & 1' & 2' & 1'''  & 2' & 1'''& 2 & 1 & 2' &1' &1\\
\hline
Z_4 &  -i & -i & + & +   & i & i & - & - & + &+&+\\
\hline
\end{array}$$
\end{table}

The most general Yukawa superpotential involving the quark and lepton fields invariant under the $Q_6 \times Z_4$ symmetry, assuming matter parity in the usual way, is:
\begin{eqnarray}
\label{WYuk}
W_{\rm Yukawa} &=& \{a_u Q_3 u_3^c H_3^u +
b_u (Q \ast H^u) u_3^c + b'_u Q_3 (H^u \ast u^c)+ c_u
(Q \star u^c) H_3^u+ u\to d \}\nonumber \\
&+& \{a_\ell L_3 e_3^c H_3^d +
b_e (L \ast H^d)  e_3^c + b'_e L_3 (H^d \ast e^c)
+ c_e (L\star  e^c) H_3^d + e\to \nu \}
\nonumber \\
&+& {M_1 \over 2} \nu^c \cdot \nu^c  + {a_{\nu^c} \over 2} \nu_3^c \nu_3^c T_3~,
\label{yukawa}
\end{eqnarray}
where we have defined
\begin{eqnarray}
x \cdot y &=& x_1 y_1+x_2y_2
~,~x \ast y =x_1 y_2+x_2y_1   ~,~x \star y=x_1 y_2-x_2y_1 ~.
\label{products}
\end{eqnarray}
We have used the explicit basis for $Q_6$ given in Ref. \cite{bk}
 and the notation $u^c \equiv (-u_1^c,~u_2^c)$ etc, for the right--handed $Q_6$ doublet fermion fields.  Note that the $Z_4$ symmetry plays no role in the construction
of Eq. (\ref{WYuk}).

\subsection{The  Higgs sector}

In order to break the $Q_6$ symmetry spontaneously while avoiding
pseudo--Nambu--Goldstone  bosons
one needs to introduce SM singlet Higgs fields.  The minimal such set
will involve a $2$, $2'$, $1'$ and two $1$'s of $Q_6$. These are listed
in Table 1.
The SM singlet $S$'s are  needed to mix
the $Q_6$ doublets $H^{u,d}$ with the $Q_6$ singlets
$H_3^{u,d}$.
Without the $Q_6$ doublet $T$ there will be  an accidental $O(2)$ symmetry
in the Higgs potential.  The $O(2)$ symmetry is violated by
the cubic coupling of $T$.  The field
$T_3$ is introduced for the Majorana mass for $\nu_3^c$,
and the $Q_6$ singlet $U$ is introduced to
generate a spontaneous CP violation and also to enable
the spontaneous breaking of $Q_6\times Z_4$ within the
SM singlet sector. Thus the SM singlet Higgs sector
employed appears to be the minimal set consistent with the demands we wish to meet.

The most general Higgs superpotential involving the Higgs fields of Table 1
invariant under the $Q_6 \times Z_4$ symmetry along with the usual matter parity (with all
the Higgs fields being even) has the form
\begin{eqnarray}
W_{\rm Higgs} &=&
W_{U}+W_{ ST}+W_{ H}~,
\label{WHiggs}
\end{eqnarray}
where
\begin{eqnarray}
W_{ U} &=&
\mu_{U}~ U^2+\lambda ~U^3+
\left(\lambda_1 ~ S_3^2 +\lambda_2 ~T_3^2+
\lambda_3 T\cdot T \right)U,
\label{WHiggsU}\\
W_{ ST} &=&
\mu_{S_3}~ S_3^2+\mu_{T}~T\cdot T +\mu_{T_3} ~T_3^2
+ \lambda'_3~T\cdot(T \otimes T)\nonumber\\
& &+\lambda'_1 [~-2 S_2 S_1 T_1+(S_1^2-S_2^2)T_2~]
+\lambda'_2 S\cdot  S T_3~,\\
\label{WHiggsST}
W_{H} &=&
\lambda''_1~H^u_3  (H^d \ast S)+
\lambda''_2~ (H^u \ast S) H^d_3 +\lambda''_3~(H^u \cdot H^d) S_3
\label{WHiggsH}
\end{eqnarray}
with the notation
\begin{eqnarray}
A\cdot(B \otimes C) &=& A_1(-B_1 C_1+B_2 C_2)+A_2(B_1C_2+B_2 C_1)~.
\label{cross}
\end{eqnarray}
Thus $T\cdot(T \otimes T)=3 T_1 T_2^2-T_1^3$.
The $Z_4$ symmetry has restricted the form of
 Eqs. (\ref{WHiggsU})-(\ref{WHiggsH}); without the $Z_4$,
the following couplings would be allowed:
\begin{equation}
W'_{\rm Higgs} =
(-H_1^u H_1^d + H_2^u H_2^d) T_1 + (H_1^u H_2^d + H_2^u H_1^d) T_2~.
\end{equation}
We wish to avoid these terms, since in their absence we can define an unbroken discrete symmetry, as discussed  below.

The Higgs potential contains $F$ terms derived from
Eqs. (\ref{WHiggsU})- (\ref{WHiggsH}),
$D$ terms associated with
$SU(2)_L \times U(1)_Y$ breaking, and the following soft SUSY breaking
Lagrangian\footnote{We have used the same symbol for the scalar components
as the superfields.}
\begin{eqnarray}
\label{Vsoft}
{\cal L}_{\rm soft} &=& m_{U}^2 |U|^2+ m_S^2(|S_1|^2+|S_2|^2)
+ m_{S_3}^2 |S_3|^2 +
m_T^2 (|T_1|^2 + |T_2|^2) + m_{T_3}^2 |T_3|^2
 \nonumber \\
 &+& m_{H_3^u}^2 |H_3^u|^2+
m_{H_3^d}^2 |H_3^d|^2 +
m_{H^u}^2(|H_1^u|^2+|H_2^u|^2) + m_{H^d}^2(|H_1^d|^2+|H_2^d|^2) \nonumber \\
& +&\left\{~B_{U}~ U^2+
B_{S_3}~ S_3^2 +B_T~T \cdot T+B_{T_3}~T_3^2
\right.\nonumber \\
&+& \left[A~U^2+A_1~ S_3^2 +
A_2~ T_3^2 +A_3~ (T \cdot T) \right]~U+ A'_3~T\cdot(T \otimes T)
\nonumber \\
&+& A'_1 [~-2 S_2 S_1 T_1+(S_1^2-S_2^2)T_2~]
+A'_2 S\cdot  S T_3 \nonumber\\
& +&\left. A''_1~H^u_3  (H^d \ast S)+
A''_2~ (H^u \ast S) H^d_3
+A''_3~(H^u \cdot H^d) S_3
+h.c.\right\}~,
\label{Lsoft}
\end{eqnarray}
where the $\cdot$ and $\ast$ products are defined in (\ref{products}).
We assume CP invariance, which implies that all the Yukawa couplings
and the parameters in the Higgs potential
are  real.
The Higgs potential would then admit two interesting minima
which leave two separate discrete symmetries ${\cal P}_I$ or ${\cal P}_{II}$ unbroken.
We analyze these two ground states in the next two sections.

\section{Ground state with unbroken interchange symmetry ${\cal P}_I$}

The following symmetry ${\cal P}_I$ is respected by the $Q_6 \times Z_4$
invariant Higgs superpotentials
Eqs. (\ref{WHiggsU}) -(\ref{WHiggsH}), and (\ref{Vsoft})
and  the $D$ terms:
\begin{eqnarray}
H_1^u \leftrightarrow H_2^u,~ H_1^d \leftrightarrow H_2^d,~S_1
 \leftrightarrow S_2,~
T_2 \rightarrow -T_2,\nonumber \\
~H_3^u \rightarrow H_3^u,~H_3^d
\rightarrow H_3^d,~S_3 \rightarrow S_3~,T_1 \rightarrow T_1,~T_3 \rightarrow T_3,~
~U \rightarrow U.
\label{PI}
\end{eqnarray}
The VEVs of the various Higgs fields can be consistently chosen such that this
symmetry remains unbroken:
\begin{eqnarray}
\label{VEV1}
&~& \left\langle H_1^{u,d} \right \rangle  = \left\langle H_2^{u,d}
 \right \rangle  = v_1^{u,d} e^{i \phi_+^{u,d}},~
\left\langle H_3^{u,d} \right \rangle =  v_3^{u,d} e^{i \phi_3^{u,d}},
 \left\langle S_1 \right \rangle  = \left\langle S_2
\right \rangle  = v_S e^{i \phi_S},
\nonumber \\
&~&~\left\langle T_1
\right \rangle  = v_T e^{i\phi_T},~ \left\langle T_2 \right \rangle  =0,~
\left\langle S_3 \right \rangle = v_{S_3} e^{i \phi_{S_3}},~
\left\langle T_3 \right \rangle = v_{T_3} e^{i \phi_{T_3}},~
\left\langle U \right \rangle = v_{U} e^{i \phi_{U}}~.
\end{eqnarray}
In Eq. (\ref{VEV1}), we have explicitly displayed the complex phases.  It should be noted that this symmetry
${\cal P}_I$ is an accidental symmetry of the Higgs potential, and is not respected by the full theory.  For
example, the Yukawa sector explicitly breaks this symmetry.
Nevertheless, the existence of ${\cal P}_I$ enables
us to choose a ground state given as in Eq. (\ref{VEV1}) consistently.

We have explicitly verified that the minimum of Eq. (\ref{VEV1}) is indeed a local minimum, and that spontaneous
breaking of $Q_6 \times Z_4$ and CP symmetries occurs without generating pseudo--Nambu--Goldstone bosons.  The scalar spectrum
of our model is in fact arrived at by meeting these requirements.

In the ground state ${\cal P}_I$, the mass matrices for the up and down quarks take the form:
\begin{eqnarray}
\label{Mud1}
M_{u,d} =\left( \matrix{0 & C_{u,d} & {B_{u,d} \over \sqrt{2}} e^{i \Delta \phi_{u,d}} \cr -C_{u,d} & 0 & {B_{u,d} \over \sqrt{2}} e^{i \Delta \phi_{u,d}}  \cr
{B'_{u,d} \over \sqrt{2}} e^{i \Delta \phi_{u,d}} & {B'_{u,d} \over \sqrt{2}} e^{i \Delta \phi_{u,d}} & A_{u,d}} \right)~.
\end{eqnarray}
Here we have defined the following parameters:
\begin{eqnarray}
\label{par}
&~& A_{u,d} = a_{u,d}~ v_3^{u,d},~B_{u,d}= \sqrt{2}~ b_{u,d}~ v_1^{u,d},~B'_{u,d} = \sqrt{2}~ b'_{u,d} ~ v_1^{u,d},~C_{u,d} = c_{u,d}~ v_3^{u,d},~ \nonumber \\
&~& \hspace*{2.0in} \Delta \phi_{u,d} = \phi_3^{u,d} - \phi_1^{u,d}~.
\end{eqnarray}
We have ignored irrelevant overall phases of the two mass matrices. CP invariance of the Lagrangian
implies that the parameters $(A_{u,d},~B_{u,d},~B'_{u,d},~C_{u,d})$ are all real.  In this case, after a common 45 degree rotation
in the (1-2) sector that would set the (1,3) and (3,1)
entries of $M_{u,d}$ of Eq. (\ref{Mud1}) to zero without inducing
CKM mixing, we can write
\begin{equation}
M_{u,d} = P_{u,d} \hat{M}_{u,d} P_{u,d}~,
\end{equation}
where $\hat{M}_{u,d}$ are real matrices given as
\begin{eqnarray}
\label{Mudhat}
\hat{M}_{u,d} = \left( \matrix{0 & C_{u,d} & 0 \cr -C_{u,d} & 0 & B_{u,d}   \cr
0 & B'_{u,d} & A_{u,d}} \right)~,
\end{eqnarray}
and $P_{u,d}$ are diagonal phase matrices given as
\begin{equation}
P_{u,d} = {\rm diag.}\{e^{-i\Delta \phi_{u,d}},~e^{i\Delta \phi_{u,d}},~1\}~.
\end{equation}
The CKM matrix is then given by
\begin{equation}
V_{\rm CKM} = O_u^T P O_d~,
\end{equation}
where $O_{u,d}$ are the orthogonal matrices that diagonalize $\hat{M}_{u,d}$ via
\begin{equation}
\label{Oud}
O_{u,d}^T \hat{M}_{u,d} M_{u,d}^T O_{u,d} = {\rm diag.} \{m^2_{u,d},~m^2_{c,s},~m^2_{t,b} \}~,
\end{equation}
and $P$ is a diagonal phase matrix
\begin{equation}
\label{P}
P = {\rm diag.} \{e^{i \phi},~ e^{-i \phi},~1\}~
\end{equation}
with $\phi = \Delta \phi_d - \Delta \phi_u$.
Since $\hat{M}_u$ and $\hat{M}_d$ each has
four real parameters, once the six quark masses are fixed, $O_u$ and $O_d$ will have
one undetermined parameter each.  These two parameters and the phase $\phi$  appearing in the
matrix $P$ of Eq. (\ref{P}) will completely fix the three CKM mixing angles and the one CP
violating phase.  That will lead to one sum rule involving the CKM mixing angles, the CP violating
phase, and the quark mass ratios.  This prediction was analyzed
in Refs. \cite{bk,Araki:2008rn}
 and shown to be fully consistent with data.

Here we update the results of Ref. \cite{bk} for the quark mixing parameter prediction.
We use the most recent values of the quark masses.  Lattice calculations have reduced the
errors in the light quark masses, which we adopt for our fits.  Furthermore, we compare the
model prediction with the global fits provided recently in Refs. \cite{Lenz:2010gu,UTfit} assuming
specific new physics contributions.  The new physics contributions are motivated by certain
discrepancies that have been observed in the CKM fits.  For example, the CP violation parameter
$\overline{\eta}$ determined from $\epsilon_K$ differs from that obtained from the decay
$B \rightarrow J/\psi K_S$ by more than 2 standard deviations.  We compare our model fits
with the best fit of the standard model, as well as with the best fit for Scenario 1 of
Refs. \cite{Lenz:2010gu,UTfit}.  This scenario is characterized by independent new contributions
$\Delta_{d,s}$ to $B_{d,s}-\overline{B}_{d,s}$ mixing amplitude.  It turns out that there is room for
small ($\sim 25\%)$ new contributions to these mixings in our model, arising from gluino--squark
box diagrams.  The $Q_6$ assignment of quarks implies that the third family squark is not
degenerate with the first two family squarks (which are nearly degenerate).  Once the quark
mass matrices are diagonalized, there will be small off--diagonal entries in the squark
mass matrix, which leads to $B_{d,s}-\overline{B}_{d,s}$ mixings.
These diagrams have been evaluated in Ref. \cite{Babu:2009nn}.  While real, these amplitudes
are still in the interesting range for new physics to influence the CKM parameter fits.  In
Ref. \cite{Kubo:2010mh} the radiative corrections to these mixing parameters, arising through
Higgs boson exchange, have been computed, and have been shown to be complex.  Thus, it appears
that the $Q_6$ model admits small deviations in the CKM fits to $B_{d,s}-\overline{B}_{d,s}$ mixings.
It should be noted, however, that the prediction of the present model agrees well with the best
fit values of the CKM fits, with or without new physics assumed.

Guided by the analytic expressions for the CKM mixing parameters
from (\ref{Mudhat}) - (\ref{P}), we have done a numerical fit to all
quark masses and mixings.  An excellent fit is obtained with the following choice of parameters at
$\mu = 1$ TeV:
\begin{eqnarray}
\label{input1}
 A_u/m_t &=&0.9963,~ B_u/m_t =0.06051,~B'_u/m_t = 0.06051,
~C_u /m_t= 1.748\times 10^{-4}, \nonumber \\
A_d/m_b &=& 0.8895,~B_d /m_b= 0.04214,
~ B'_d/m_b =0.4554,~ C_d /m_b=-5.043\times 10^{-3},\nonumber\\
\phi  &= &0.71875.
\end{eqnarray}
 The resulting mass eigenvalues at $\mu = 1$ TeV are:
\begin{eqnarray}
&~& m_u = 1.25 ~{\rm MeV},~ m_c = 552~ {\rm MeV},\nonumber \\
&~& m_d = 2.74 ~{\rm MeV},~ m_s =50.0 ~{\rm MeV},
\label{masses1}
\end{eqnarray}
where we have used $m_t = 150.3 ~{\rm GeV}$ and
$m_b = 2.46 ~{\rm GeV}$.
These values are to be compared with quark masses extrapolated
from low energy scale to $\mu = 1$ TeV \cite{Xing:2007fb}:
\begin{eqnarray}
&&m_u=0.85 \sim 1.65\ {\rm MeV}~,~
m_d=2.05\sim 2.90\ {\rm MeV}\ ,\nonumber\\
&&m_s=39.6\sim 64.4\ {\rm MeV}~,~
m_c=502\sim 570\ {\rm MeV}\ ,\nonumber\\
&&m_b=2.39\sim 2.53\ {\rm GeV}~,~
m_t=148.9\sim 151.6\ {\rm GeV}~,
\label{masses0}
\end{eqnarray}
where we have updated the result of \cite{Xing:2007fb}
by using the updated quark masses given in PDG 2011 \cite{Nakamura:2010zzi},
while neglecting the uncertainties due to the RG running.
The input values of Eq. (\ref{input1}) give also the following output for the
CKM parameters:
\begin{eqnarray}
  \lambda & =& 0.2252,~A=0.7962~,~
\bar{\rho}=0.1613~,~ \bar{\eta}=0.4230~,\nonumber\\
  \sin 2\beta &=& 0.8042~,~
\alpha=84.1 ~\mbox{[deg]}~,~
  \beta=26.8~\mbox{[deg]}~,~
  \gamma=~69.1\mbox{[deg]}~,
\end{eqnarray}
which should be compared with the fit result of
the CKMfitter group (scenario I) \cite{Lenz:2010gu}
\begin{eqnarray}
\lambda &=& 0.22542\pm 0.00077~,~
A=0.801^{+0.024}_{-0.017}~,\label{lambda}\\
\bar{\rho} &=& 0.159^{+0.036}_{-0.035}~,~
\bar{\eta}=0.438^{+0.019}_{-0.029}~,~
\sin 2\beta=0.813^{+0.022}_{-0.068}~,\nonumber\\
\alpha &=& 79^{+22}_{-15}~\mbox{[deg]}~,~
\beta=27.2^{+1.1}_{-3.1}~\mbox{[deg]}~,~
\gamma=70.0^{+4.3}_{-4.5}~\mbox{[deg]}~.
\label{ckmfitter}
\end{eqnarray}

\begin{figure}[ht]
\hspace{2cm}
\includegraphics*[width=0.8\textwidth]{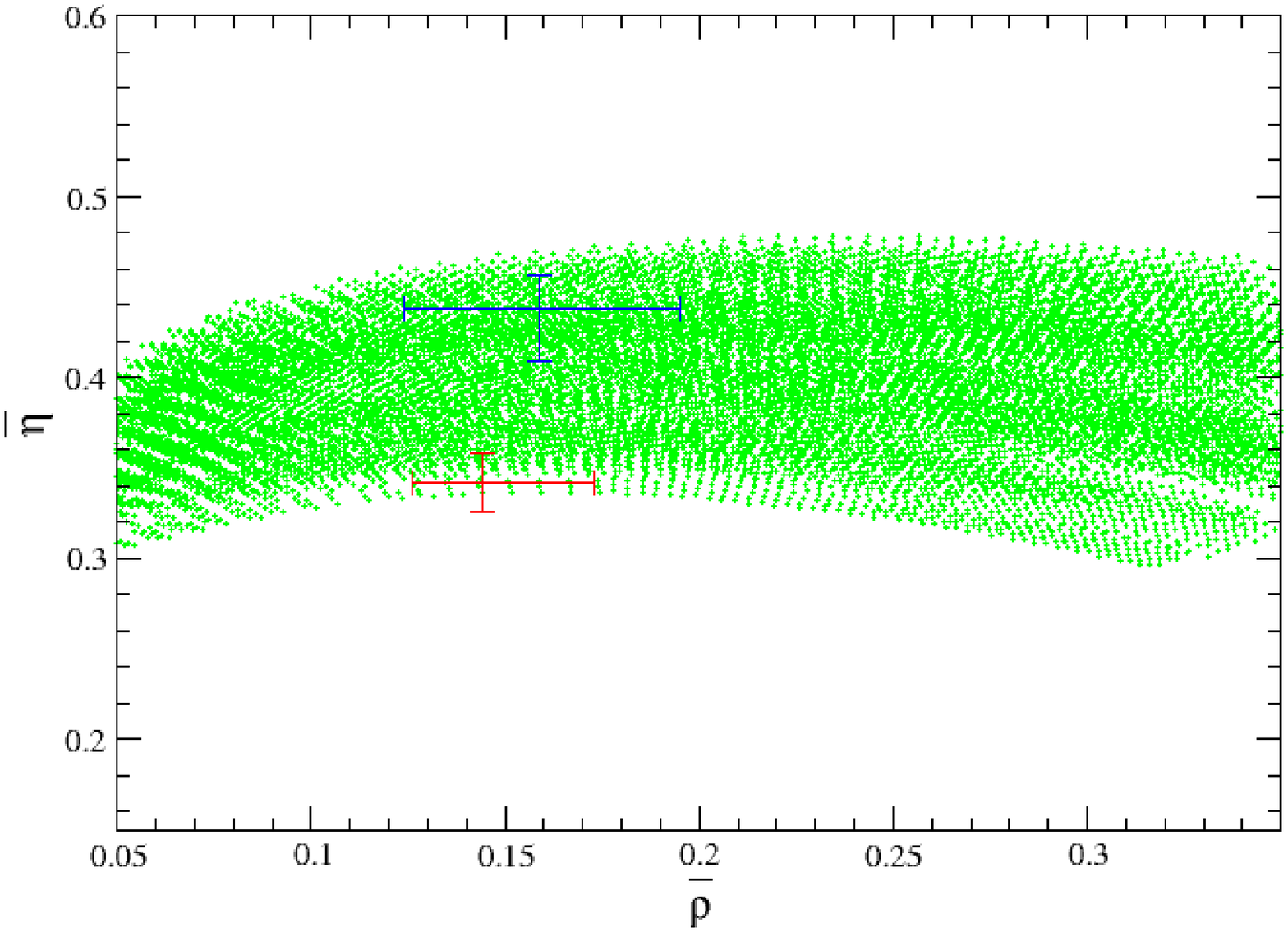}
\caption{\label{prediction1}\footnotesize
 The prediction in the
$\bar{\rho}-\bar{\eta}$ plane for the model ${\cal P}_{I}$, where we have used
as the input parameters; the quark masses,
$\lambda$ and $A$ given in Eqs. (\ref{masses0}) and (\ref{lambda}), respectively.
We also have imposed the constraints on the quark mass ratios\cite{Nakamura:2010zzi}:
$2m_s/(m_u+m_d)=22\sim 30,~m_s/m_d=17\sim 22,
m_u/m_d=0.35\sim 0.60$.
The crosses  are the CKMfitter group values \cite{Lenz:2010gu};
 blue (scenario I) and red (SM).}
\end{figure}

\begin{figure}[ht]
\hspace{2cm}
\includegraphics*[width=0.8\textwidth]{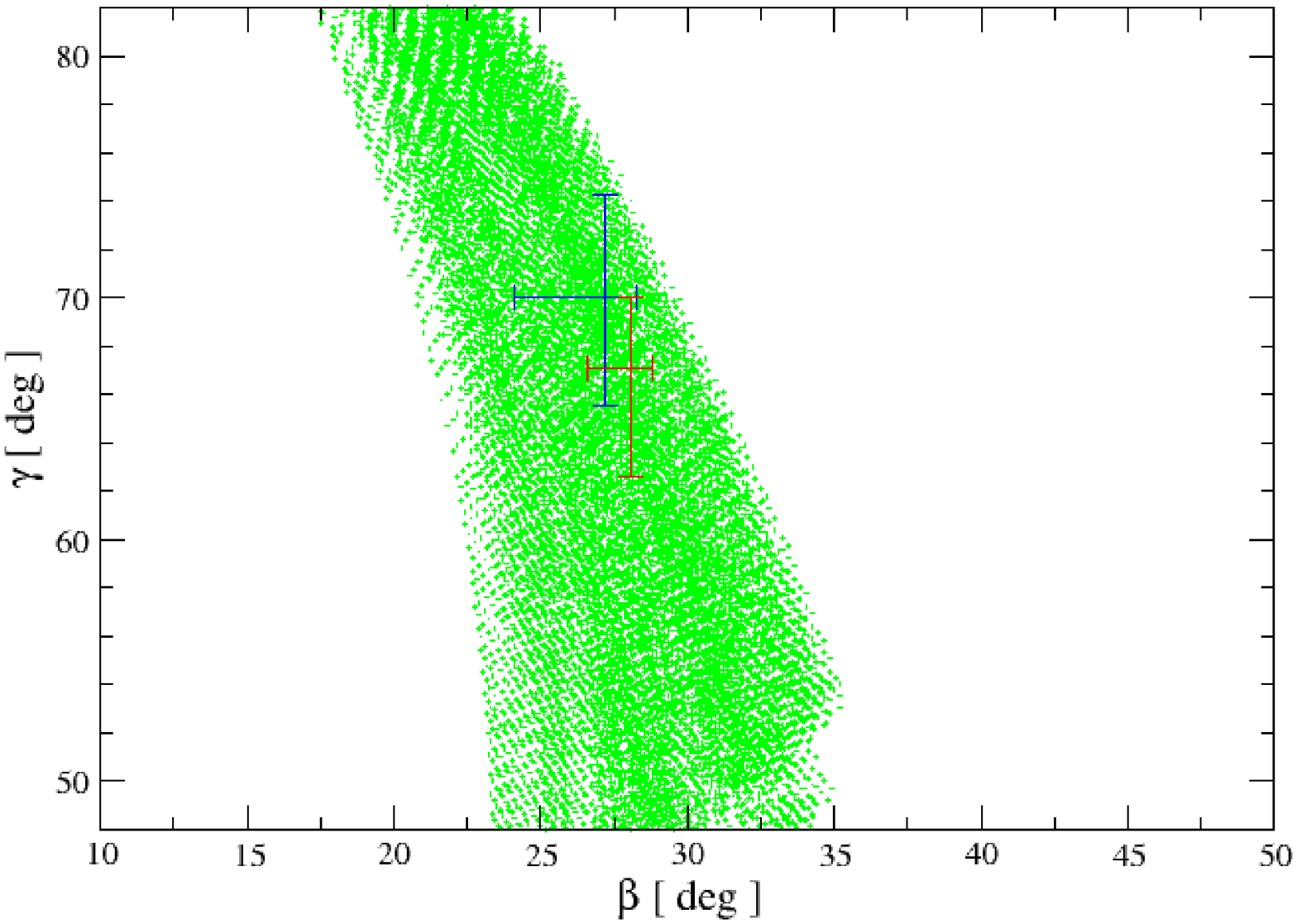}
\caption{\label{prediction2}\footnotesize
 The prediction in the
$\beta-\gamma$ plane for the model ${\cal P}_{I}$.
The input parameters and the constraints are the same as
for Fig.~\ref{prediction1}.}
\end{figure}

Since there are nine model parameters for six quark masses
and four CKM mixing parameters, we can make one prediction
in a two dimensional plane if we fix  eight of the nine model parameters.
To fix these eight parameters we use the quark masses,
$\lambda$ and $A$ given in (\ref{masses0}) and (\ref{lambda}), respectively.
Fig.~\ref{prediction1} shows the prediction in the
$\bar{\rho}-\bar{\eta}$ plane, and Fig.~\ref{prediction2}
shows the prediction in the
$\beta-\gamma$ plane.
 The CKMfitter group best fit values (\ref{ckmfitter})
are also indicated in these figures.
We see from  Eqs. (\ref{masses1}),
Fig.~\ref{prediction1} and Fig.~\ref{prediction2} that
 the model ${\cal P}_{I}$ reproduces the quark masses, CKM mixings and the CP violating
phase in an excellent way.

\section{New ground state with unbroken interchange symmetry ${\cal P}_{II}$}

The same Higgs potential as derived from Eqs. (\ref{WHiggs}),
and the soft SUSY breaking Lagrangian (\ref{Vsoft}) including the $D$ terms, with all
parameters taken to be real so that CP is an exact symmetry,
admits a new unbroken interchange symmetry as given below:
\begin{eqnarray}
H_1^u \leftrightarrow H_2^{u *},~H_1^d \leftrightarrow H_2^{d *},~
S_1 \leftrightarrow S_2^*,~T_2 \rightarrow -T_2^*,\nonumber \\
H_3^u \rightarrow H_3^{u *},~
H_3^d \rightarrow H_3^{d *},~S_3 \rightarrow S_3^*,~
~T_1 \rightarrow T_1^*,~T_3 \rightarrow T_3^*,
~U \rightarrow U^*.
\label{PII}
\end{eqnarray}
This symmetry ${\cal P}_{II}$ enables us to choose a ground state given by
\begin{eqnarray}
\label{VEV2}
&~& \left\langle H_1^u \right \rangle = v_1^u e^{-i\phi_u},~
\left\langle  H_2^{u} \right \rangle = v_1^u e^{i\phi_u},~
\left\langle H_1^d \right \rangle = v_1^d e^{-i\phi_d},~
\left\langle  H_2^{d} \right \rangle = v_1^d e^{i\phi_d},~\nonumber \\
& &\left\langle H_3^u\right\rangle  = v_3^u,~
\left\langle H_3^d\right\rangle  = v_3^d,~
\nonumber \\
&~& \left\langle S_1 \right\rangle = v_S e^{-i \phi_S},~
\left\langle S_2 \right\rangle = v_S e^{i \phi_S},~
\left\langle S_3 \right\rangle = v_{S_3},~\\
& &\left\langle T_1 \right\rangle =  v_{T_1},~
 \left\langle T_2 \right\rangle = -iv_{T_2},~
  \left\langle T_3 \right\rangle = v_{T_3},~
    \left\langle U \right\rangle = v_{U},
  \nonumber
\end{eqnarray}
where the complex phases are all explicitly displayed.
Note that there are only three phases, $\phi_S$, $\phi_u$ and
$\phi_d$ in the VEVs, along with a purely imaginary VEV of $T_2$.

In the background ${\cal P}_{II}$,
the fermion mass matrices $M_{u,d}$ following from Eq. (\ref{WYuk}) take the form
\begin{eqnarray}
\label{Mud2}
M_{u,d} = \left( \matrix{0 & C_{u,d} & {B_{u,d} \over \sqrt{2}} e^{-i\phi_{u,d}} \cr -C_{u,d} & 0 & {B_{u,d} \over \sqrt{2}} e^{i \phi_{u,d}}  \cr
{B'_{u,d} \over \sqrt{2}} e^{-i \phi_{u,d}} & {B'_{u,d} \over \sqrt{2}} e^{i  \phi_{u,d}} & A_{u,d}} \right)
\end{eqnarray}
with the parameters as defined in Eq. (\ref{par}). CP invariance of the Lagrangian implies that the parameters $\{A_{u,d},~B_{u,d},~B'_{u,d},~C_{u,d}\}$ are all real.

Model ${\cal P}_{II}$, while different from model ${\cal P}_I$, is just as predictive in the quark sector as
 ${\cal P}_I$.  It is then interesting to see if the quark mixing sum rule of ${\cal P}_{II}$ is consistent with
 data.  To address this question we proceed to diagonalize $M_{u,d}$ of Eq. (\ref{Mud2}).
The phases in the matrices of Eq. (\ref{Mud2}) can be factorized:
\begin{equation}
M_{u,d} = P_{u,d} M^{r}_{u,d} P_{u,d}
\end{equation}
where
\begin{equation}
P_{u,d} = {\rm diag.}\{e^{i \phi_{u,d}},~e^{-i \phi_{u,d}},~1 \}
\end{equation}
with $M^r_{u,d}$ given as in Eq. (\ref{Mud2}), but with $\phi_{u,d}$ set to zero.  Quark field redefinitions
can absorb the phases in $P_{u,d}$, however a phase matrix will then appear in the quark mixing matrix:
\begin{equation}
P = {\rm diag.} \{ e^{i \phi},~e^{-i\phi},~1 \}~,
\end{equation}
where
\begin{equation}
\phi = \phi_d-\phi_u~.
\end{equation}
Now we do a 45 degrees rotation in the (1-2) plane to bring $M^r_{u,d}$ into $\hat{M}_{u,d}$ as given in
Eq. (\ref{Mudhat}), but this will generate a non-trivial quark mixing matrix given by
\begin{eqnarray}
\label{K}
K = \left(\matrix{\cos\phi & i \sin\phi & 0 \cr i \sin\phi & \cos \phi & 0 \cr 0 & 0 & 1}\right)~.
\end{eqnarray}
The CKM mixing matrix is then obtained as
\begin{equation}
\label{V}
V_{\rm CKM} = O_u^T K O_d~,
\end{equation}
where $O_{u,d}$ diagonalize the matrices of Eq. (\ref{Mudhat}) as specified in Eq. (\ref{Oud}).

Using the approximate analytic expressions for the CKM mixing parameters, we have done a numerical fit to all
quark masses and mixings within this model.  An excellent fit is obtained with the following choice of parameters at
$\mu = 1$ TeV:
\begin{eqnarray}
\label{input2}
 A_u/m_t &=&0.01389,~ B_u/m_t =-0.003282,~B'_u/m_t = 0.9999,
~C_u /m_t= 1.381\times 10^{-3}, \nonumber \\
A_d/m_b &=& 0.9020,~B_d /m_b= 0.04512,
~ B'_d/m_b =0.4297,~ C_d /m_b=4.554\times 10^{-3},\nonumber\\
\phi  &= &0.1038.
\end{eqnarray}
 The resulting mass eigenvalues at $\mu = 1$ TeV are:
\begin{eqnarray}
&~& m_u = 1.12~{\rm MeV},~ m_c = 535~ {\rm MeV},\nonumber \\
&~& m_d = 2.27 ~{\rm MeV},~ m_s =50.0 ~{\rm MeV},
\label{masses1new}
\end{eqnarray}
where we have used $m_t = 150.3 ~{\rm GeV}$ and
$m_b = 2.46 ~{\rm GeV}$ as in the case of ${\cal P}_I$.
These values are to be compared with quark masses
given in Eq. (\ref{masses0}).
The input values of Eq. (\ref{input2}) give the  output for the
CKM parameters:
\begin{eqnarray}
  \lambda & =& 0.2254,,~A=0.7987~,~
\bar{\rho}=0.1575~,~ \bar{\eta}=0.4231~,\nonumber\\
  \sin 2\beta &=& 0.8021~,~
\alpha=83.7~\mbox{[deg]}~,~
  \beta=26.7~\mbox{[deg]}~,~
  \gamma=69.9\mbox{[deg]}~,
\end{eqnarray}
which should be compared with the fit result of the CKMfitter group
(\ref{ckmfitter}).

\begin{figure}[ht]
\hspace{2cm}
\includegraphics*[width=0.8\textwidth]{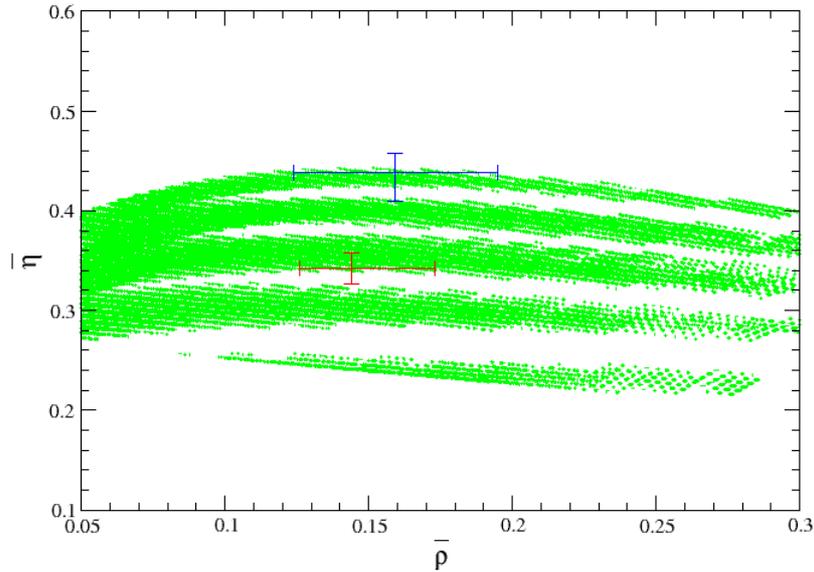}
\caption{\label{prediction3}\footnotesize
 The prediction in the
$\bar{\rho}-\bar{\eta}$ plane for the model ${\cal P}_{II}$, where we have used
as the input parameters; the quark masses,
$\lambda$ and $A$ given in Eqs. (\ref{masses0}) and (\ref{lambda}), respectively.
We also have imposed the constraints on the quark mass ratios\cite{Nakamura:2010zzi}:
$2m_s/(m_u+m_d)=22\sim 30,~m_s/m_d=17\sim 22,
m_u/m_d=0.35\sim 0.60$.
The crosses  are the CKMfitter group values \cite{Lenz:2010gu};
 blue (scenario I) and red (SM).}
\end{figure}

\begin{figure}[ht]
\hspace{2cm}
\includegraphics*[width=0.8\textwidth]{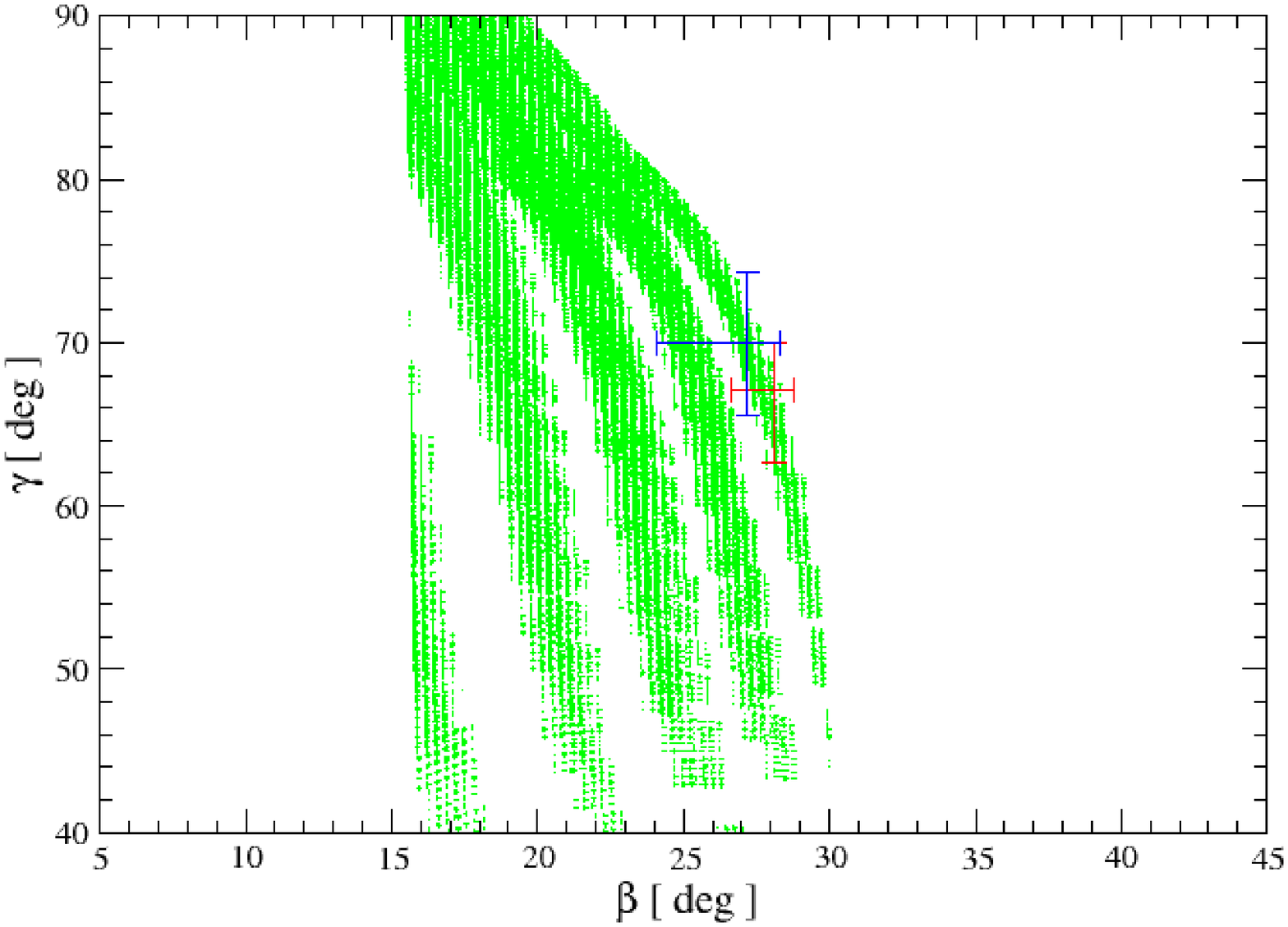}
\caption{\label{prediction4}\footnotesize
 The prediction in the
$\beta-\gamma$ plane for the model ${\cal P}_{II}$.
The input parameters and the constraints are the same
as for Fig.~\ref{prediction3}.}
\end{figure}

Fig.~\ref{prediction3} shows the prediction in the
$\bar{\rho}-\bar{\eta}$ plane, and Fig.~\ref{prediction4}
shows the prediction in the
$\beta-\gamma$ plane for model ${\cal P}_{II}$.
The CKMfitter group best fit values (\ref{ckmfitter})
as well as the SM best fit values
are indicated in these plots.
As in the case of ${\cal P}_I$,
we see from  Eqs. (\ref{masses1new}),
Fig.~\ref{prediction3} and Fig.~\ref{prediction4} that
the model ${\cal P}_{II}$ also reproduces the quark masses, CKM mixings and the CP violating
phase in an excellent way.


\section{Predictive scenario for neutrino mixing}

The lepton sector of model ${\cal P}_I$  with the $Q_6$ assignment given in
Table 1 has been studied in Ref. \cite{bk}, and therefore
we will not discuss it further here.
It is interesting to see if there are any constraints on neutrino oscillation
parameters for model  ${\cal P}_{II}$.
Here we explore an alternative possibility of the $Q_6$ assignment for the leptons,
which is given in Table 2.

\begin{table}[h]
\caption{An alternative $Q_6 \times Z_4$ assignment for the leptons.}
\vspace*{-0.05in}
$$\begin{array}{|c||c|c|c|c|c|}
\hline
~ &  L & \{ e^c,\nu^c \} &  L_3 & e_3^c   & \nu_3^c  \\
\hline
Q_6 & 2' & 2'& 1 & 1 & 1''  \\
\hline
Z_4 &  -i & +&-i&+  &+  \\
\hline
\end{array}$$
\end{table}

In this new scenario, the leptonic part of the superpotential (\ref{WYuk}) becomes
\begin{eqnarray}
\label{WYuk-l}
W_{\rm Yukawa}\ell &=&
 b_e (L \cdot H^d) e_3^c + b'_e L_3( H^d \cdot e^c)
 + c_e (L \otimes e^c)\cdot H^d  \nonumber \\
&+& a_\nu L_3 \nu_3^c H_3^u + b'_\nu L_3 (H^u \cdot \nu^c)
+ c_\nu (L\otimes  \nu^c)\cdot H^u \nonumber \\
&+& {M_1 \over 2} \nu^c \cdot \nu^c
+ {a_{\nu^c} \over 2} \nu_3^c \nu_3^c T_3~,
\end{eqnarray}
where the $\cdot$  and $\otimes$ products are defined in (\ref{products}) and
 (\ref{cross}), respectively.

The Majorana
mass matrix for the right--handed neutrinos is given by
\begin{eqnarray}
\label{Maj}
M_{\nu^c} = \left(\matrix{M_1 & 0 & 0 \cr 0 & M_1 & 0 \cr 0 & 0 & M_3}  \right)~,
\end{eqnarray}
where $M_3 = a_{\nu^c} ~v_{T_3}$.  Note that $M_1$ and $M_3$ are both real.
The Dirac neutrino and charged lepton mass matrices are:
\begin{eqnarray}
M_{\nu^D} &=& \left( \matrix{-C_{\nu}e^{i \phi_u} & C_{\nu}e^{-i \phi_u} & 0 \cr
 C_{\nu}e^{-i \phi_u} &  C_{\nu}e^{i \phi_u} & 0   \cr
B'_{\nu}e^{i \phi_u} & B'_{\nu}e^{-i \phi_u} & A_{\nu}} \right)~,~
M_{\ell} = \left( \matrix{-C_{\ell}e^{i \phi_d} & C_{\ell}e^{-i \phi_d}
 & B_{\ell}e^{i \phi_d} \cr C_{\ell}e^{-i \phi_d}
& C_{\ell}e^{i \phi_d} & B_{\ell}e^{-i \phi_d}   \cr
 B'_{\ell}e^{i \phi_d} & B'_{\ell}e^{-i \phi_d} & 0} \right)~.
 \label{Mlep2}
\end{eqnarray}
The light neutrino Majorana mass matrix is found (by the seesaw formula) to be
\begin{eqnarray}
\label{Mnulight2}
M_\nu^{\rm light} = m_0 \left(\matrix{ 2 \rho^2_2 \cos (2\phi_u) & 0
& -2 i \rho_2\rho_4 \sin (2\phi_u) \cr
0  & 2 \rho^2_2 \cos (2\phi_u) & 2 \rho_2\rho_4 \cr
 -2 i \rho_2\rho_4 \sin (2\phi_u)  & 2 \rho_2\rho_4  & -\rho_3^2+2 \rho^2_4 \cos (2\phi_u)
}\right)~,
\end{eqnarray}
where
\begin{eqnarray}
\rho_2^2 &=& (C_{\nu})^2 / M_1~,~
\rho_3^2 = -(A_{\nu})^2/M_3~,~\rho_4^2 =(B'_{\nu})^2/M_1~.
\end{eqnarray}
We have assumed that $ M_1$ is positive, while $M_3$ is negative.
When $\phi_u=0$, the neutrino mass matrix is exactly the same as the matrix
discussed  in \cite{Kubo:2003iw}, and yields only a tiny
$U_{e3} \sim m_e/m_\mu \sim 10^{-3}$,
where $U_{e3}$ is the $(e,3)$ element of the Maki-Nakagawa-Sakata
(MNS) mixing matrix $U_{\rm MNS}$.
It was also shown there that the mass matrix (\ref{Mnulight2}) with $\phi_u=0$ can yield
consistent neutrino masses and mixing only if $M_3$ is negative, and
the mass spectrum is inverted.
This conclusion also applies to the present case with non-vanishing $\phi_u$.
For non-zero $\phi_u$, we obtain $U_{e3} \sim \sin 2\phi_u$, which can be small
or large.  We vary $|U_{e3}|$ in its entire range allowed by experiments and correlate
its value with other observables.

\begin{figure}[htb]
\hspace{1cm}
\includegraphics*[width=0.8\textwidth]{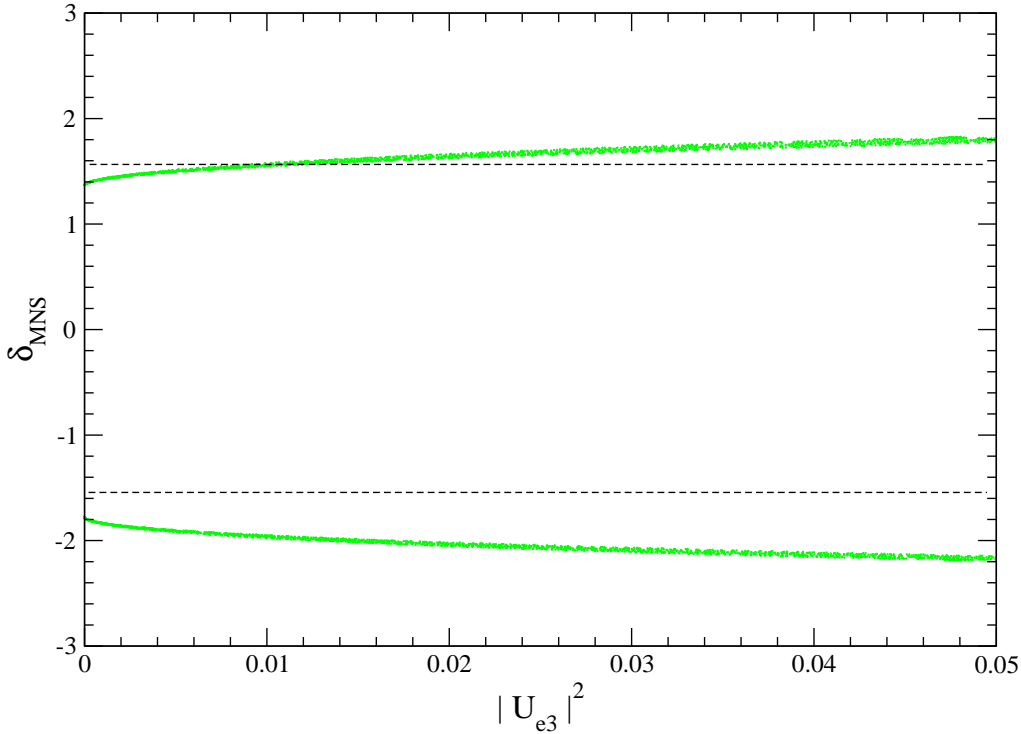}
\caption{\label{prediction-n-1}\footnotesize
 The prediction in the
$|U_{e3}|^2-\delta_{\rm MNS}$ plane for the model ${\cal P}_{II}$
with the $Q_6$ assignment of the leptons given in Table 2,
where we have used the parameters given in (\ref{input4}),
and $\phi_d=\phi_u+0.1038$.
The dashed vertical line is the
maximal CP violation.
}
\end{figure}

\begin{figure}[htb]
\hspace{1cm}
\includegraphics*[width=0.8\textwidth]{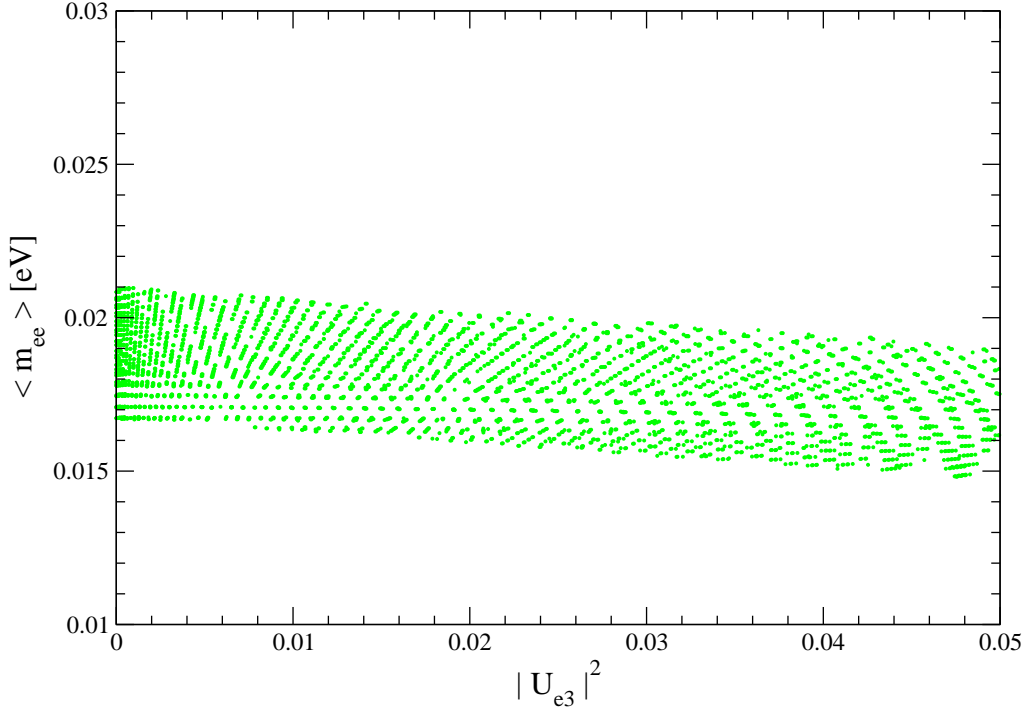}
\caption{\label{prediction-n-2}\footnotesize
 The prediction in the
$|U_{e3}|^2-<m_{ee}>$
 plane for the same input parameters as
Fig.~\ref{prediction-n-1}.
}
\end{figure}

\begin{figure}[htb]
\hspace{1cm}
\includegraphics*[width=0.8\textwidth]{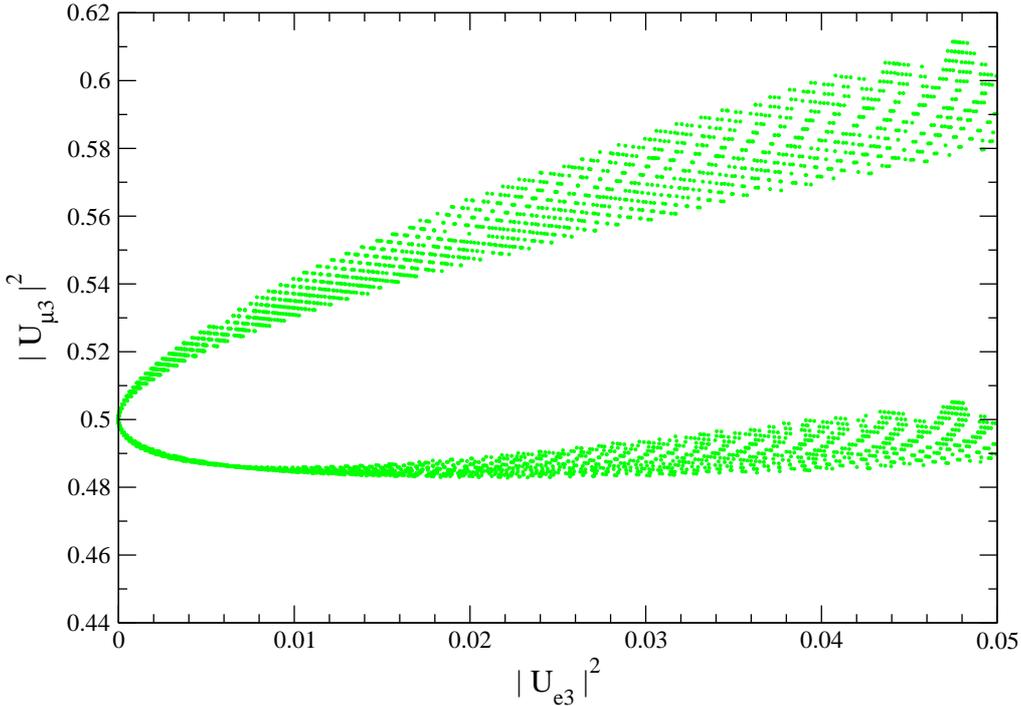}
\caption{\label{prediction-n-3}\footnotesize
$|U_{\mu3}|^2$ against $|U_{e3}|^2$
for the same input parameters as Fig.~\ref{prediction-n-1}. }
\end{figure}

We make the matrix (\ref{Mnulight2})  real by redefining $\nu_1=i \nu_1'$.
The resulting mass matrix $\hat{ M}_\nu^{\rm light}$ can
be diagonalized by an orthogonal matrix ${\cal O}_\nu$ as
${\cal O}_\nu^T \hat{ M}_\nu^{\rm light} {\cal O}_\nu$.
As for the charged lepton mass matrix $M_{\ell}$, we can obtain  hierarchical masses, e.g.,
$m_e \sim B'_{\ell}~,~m_\mu \sim C_{\ell}~,~
m_\tau \sim B_{\ell}$. Keeping this in mind we rotate $M_{\ell}$ according to
\begin{eqnarray}
\hat{M_{\ell}} &=& P_L M_{\ell} P_R ~,
\end{eqnarray}
where
\begin{eqnarray}
P_L &=&
 \frac{1}{\sqrt{2}}\left( \matrix{e^{-i \phi_d} & -e^{i \phi_d} & 0 \cr
-i e^{-i \phi_d} & -i e^{i \phi_d} & 0   \cr
0 & 0 & \sqrt{2}} \right)~,~
P_R=\frac{1}{\sqrt{2}}\left( \matrix{e^{-i \phi_d} & e^{-i \phi_d} & 0 \cr
-e^{i \phi_d} &  e^{i \phi_d} & 0   \cr
0 & 0 & \sqrt{2}} \right)~.
\label{pLpR}
\end{eqnarray}
Then we consider $\hat{M_{\ell}}\hat{M_{\ell}}^\dag$ in the limit
$B'_{\ell}\rightarrow 0$, (i.e. $m_e \rightarrow 0$), and find
\begin{eqnarray}
\hat{M_{\ell}}\hat{M_{\ell}}^\dag
&=&
\left( \matrix{C_{\ell}^2 (3-\cos(4 \phi_d)) & C_{\ell}^2\sin(4 \phi_d) & 0\cr
 C_{\ell}^2\sin(4 \phi_d)& 2 B_{\ell}^2+C_{\ell}^2 (1+\cos(4 \phi_d)  & 0   \cr
0 & 0 &0} \right)~.
\label{Mlep3}
\end{eqnarray}
The eigenvalues in the limit are
\begin{eqnarray}
m_e^2 &=&0~,~m_\mu^2 \simeq C_{\ell}^2 (3-\cos(4 \phi_d))~,~
m_\tau^2\simeq 2 B_{\ell}^2+C_{\ell}^2 (1+\cos(4 \phi_d)~,
\end{eqnarray}
and the (inverse) diagonalizing orthogonal matrix
(${\cal O}_{\ell}^T\hat{M_{\ell}}\hat{M_{\ell}}^\dag{\cal O}_{\ell}$) is found to be
\begin{eqnarray}
{\cal O}_{\ell}^T &\simeq&\left( \matrix{0 & 0& 1\cr
1& -\frac{C_{\ell}^2}{2B_{\ell}^2}\sin(4\phi_d) & 0   \cr
\frac{C_{\ell}^2}{2B_{\ell}^2}\sin(4\phi_d) & 1 &0} \right)~.
\label{Olep}
\end{eqnarray}
Since the relative phase $\phi=\phi_d-\phi_u$ is fixed in the quark sector,
there are seven independent parameters in the lepton sector.
We use \cite{Schwetz:2008er}:
\begin{eqnarray}
m_e &=& 0.511~ \mbox{MeV}~,~m_\mu = 105.7~ \mbox{MeV}~,~
m_\tau = 1.777~ \mbox{GeV}~,~|U_{e2}|^2=0.318~{}^{+0.019}_{-0.016}
\nonumber\\
\Delta m_{13}^2 &=&(2.40~{}^{+0.12}_{-0.11}  )\times10^{-3}
~\mbox{eV}^2~,~
\Delta m_{21}^2 =(7.59~{}^{+0.23}_{-0.18}  ) \times 10^{-5}
~\mbox{eV}^2
\label{input4}
\end{eqnarray}
as input parameters.
The MNS neutrino mixing matrix is then given by
\begin{eqnarray}
U_{\rm MNS} &=&{\cal O}_{\ell}^T P_L  P_\nu {\cal O}_\nu
\times {\rm diag.} \{~1~,~i~,~1~\}~,
\end{eqnarray}
where
the last phase factor multiplied with $U_{\rm MNS}$
is the Majorana phase, and $P_\nu={\rm diag.}\{ ~i~,~1~,~1~ \}$,
which was introduced to make the matrix (\ref{Mnulight2}) real.
In the lepton sector we have only one  free
phase  $\phi_u$, which controls
$U_{e3}$. In the following calculations we
use $\phi_d=\phi_u+0.1038$ (see (\ref{input2}).

Fig.~\ref{prediction-n-1} shows the Dirac phase $\delta_{\rm MNS}$ (in the convention
of Ref. \cite{Nakamura:2010zzi}) against $|U_{e3}|^2$.
We see that the model predicts
 nearly maximal CP violation.
This can be understood as follows. Consider
the limit $m_e,\phi_u,\phi_d \rightarrow 0$.
In this limit, only $P_\nu$ contributes to $\delta_{\rm MNS}$,
and the first element of $P_\nu$, $e^{i \pi/2}$, appears
as the Dirac phase.

It is possible to predict the
effective neutrino mass $<m_{ee}>
=| m_{\nu_1} U_{e1}^2+m_{\nu_2} U_{e2}^2+
m_{\nu_3} U_{e3}^2|$ for  neutrinoless double beta decay
as a function of $|U_{e3}|$.
Note that the first row of ${\cal O}_{\ell}^T P_L  P_\nu $ is
${\rm diag.}\{~0~,~0~,~1~\}$ in the $m_e \rightarrow 0$ limit.
Since ${\cal O}_{\nu}$ is real,
the first and third elements of the first row of $U_{\rm MNS}$
are real, while the second element is purely imaginary.
Therefore,
\begin{eqnarray}
<m_{ee}> &\simeq & |m_{\nu_1} \cos^2 \theta_{\rm sol}
-m_{\nu_2} \sin^2\theta_{\rm sol}| \simeq m_{\nu_2}\cos 2\theta_{\rm sol}
\simeq 0.4 ~m_{\nu_2}.
\end{eqnarray}
In Fig.~\ref{prediction-n-2} we plot the prediction in the
$|U_{e3}|^2-<m_{ee}>$ plane, which
verifies the rough estimate above.
The main contribution to $|U_{\mu 3}|$ comes from ${\cal O}_{\ell}$.
In the limit $m_e,  \phi_u \rightarrow 0$,
 it is exactly $1/\sqrt{2}$, so the maximal mixing.
The deviation from the maximal mixing
has terms proportional to $m_e/m_\mu$ and to
 $\sin 2 \phi_u$. In Fig.~\ref{prediction-n-3} we plot $|U_{\mu 3}|^2$ against
 $|U_{e 3}|^2$,
verifying our expectation.  Note that the entire range of $|U_{e3}|$ allowed
by experiments currently is also allowed by atmospheric neutrino oscillations.
But once the $|U_{e3}|$ is measured, the model will make precise prediction
for $|U_{\mu3}|$ which can be scrutinized with improved precision experiments.

\section{Conclusions}

The $Q_6$ model of flavor is constructed to solve the SUSY flavor problem
of the supersymmetric standard model.  It also yields an interesting prediction
for the quark mixing parameters, which compares very well with experimental data.
An unbroken interchange symmetry plays an important role in obtaining the
quark mixing parameter prediction.  In this paper we have updated this prediction,
and compared it with the best fit values within the standard model as well as
with new physics contributions assumed in $B_{d,s}-\overline{B}_{d,s}$ mixing
amplitudes.  The model prediction is in very good agreement with the data.

A major observation of the present paper is the existence of a new minimum
that violates CP symmetry spontaneously, but leaves a new interchange symmetry
unbroken.  In this minimum, there is again a prediction for quark mixing parameters.
We have analyzed this prediction and found that it fits data (within the CKM model
and with new physics included) rather well.  We have extended this symmetry to
the leptonic sector, and have found various correlations between neutrino oscillation
parameters.  

We conclude with several comments on the new solution found.

{\bf (1)} The SUSY flavor problem is solved in the new ground state ${\cal P}_{II}$ in the same
way it is solved in ${\cal P}_I$.  $Q_6$ invariance requires the first two family squarks and sleptons
to be degenerate in mass, which provides the needed SUSY GIM mechanism.
Since after $Q_6$ breaking the
$Q_6$ doublet and singlet quark states mix,
there is residual flavor violation mediated by the SUSY particles,
but such FCNC processes are well within experimental limits.

{\bf (2)} The SUSY CP problem is solved in the model by virtue of spontaneous CP violation.
The fundamental parameters in the Lagrangian are all real, complex phases develop only spontaneously
via the VEVs of $H^{u,d}$ and $S,T, U$ fields.  This implies that the soft SUSY breaking parameters
such as the gluino mass are all real, which alleviates bulk of the SUSY phase problem.  The trilinear
SUSY breaking $A$--terms are not proportional to the corresponding Yukawa couplings, however the
phases in these $A$--terms, since they arise spontaneously, will align with the phases in the fermion
mass matrices.  Thus the $A$--terms do not generate CP violation.  There is CP violation arising from
the $\mu$--terms, but as suggested in Ref. \cite{Babu:2009nn}, if the Higgsino masses are parametrically smaller than
the squark and slepton masses, this CP violation is not excessive.  We also note that in the new
minimum ${\cal P}_{II}$, the spontaneously induced phase that is necessary for KM CP violation is
rather small, $\sim 0.1038$.
One can then assume an approximate CP symmetry for the entire Lagrangian,
where all the phases remain small, of this order.  This will further suppress the SUSY phase effects.

{\bf (3)}  The new interchange symmetry ${\cal P}_{II}$ might appear to be CP transformation,
but it actually is not.  If it were CP transformation, when extended to the fermion Yukawa sector, that
would make the parameters $c_{u,d,\ell,\nu}$ in Eq. (\ref{WYuk}) purely imaginary.  CP violation will
then disappear from the CKM matrix, as it should, since this symmetry remains unbroken.  The symmetry
${\cal P}_{II}$ is an accidental symmetry of the Higgs potential, and is not respected by the Yukawa
couplings, just as it was for the interchange symmetry ${\cal P}_I$.
 This state leads to a new sum rule involving the quark masses and
CKM mixing parameters, which is found to be in good agreement with data.  Extension of the
model to the neutrino sector, by changing the $Q_6$ assignment of the leptons,
can lead to a predictive scenario.  In this version we find
that neutrino mass hierarchy is inverted
with nearly maximal CP violation along with
nearly maximal mixing of atmospheric neutrinos.  Thus the model lends itself to experimental scrutiny
in the near future.

{\bf (4)} The question of whether it is possible to obtain
a large CP violation  in the $B^0_s-\bar{B}^0_s$ mixing for the case of ${\cal P}_{II}$,
as in the case of ${\cal P}_{I}$ \cite{Kubo:2010mh},
remains to be studied.
To distinguish two ground states of the same model,
precise measurements of the CKM parameters \cite{Aushev:2010bq}
and  neutrino oscillation parameters  \cite{Ichikawa:2010zza} as well as
precise determination of the quark masses
\cite{Colangelo:2010et} are indispensable.

\section*{Acknowledgments}  This work was initiated during the YITP workshop
``Summer Institute 2009" (YITP-W-09-08) in Fujiyoshida.
The authors wish to thank the Yukawa Institute for Theoretical Physics at
Kyoto University for hospitality.  The work of KSB is supported in part by the US Department of Energy Grant Nos.
DE-FG02-04ER41306 and DE-FG02-ER46140. The work of JK is partially supported by a Grant-in-Aid for
Scientific Research (C) from Japan Society for the Promotion of Science (No. 22540271).


\begin{thebibliography}{99}


\bibitem{pakvasa}
  S.~Pakvasa and H.~Sugawara,
  Phys.\ Lett.\  B {\bf 73} (1978) 61;
  T.~Brown, N.~Deshpande, S.~Pakvasa and H.~Sugawara,
  Phys.\ Lett.\  B {\bf 141} (1984) 95;
  E.~Ma,
  Phys.\ Rev.\  D {\bf 43} (1991) 2761;
  E.~Ma,
  Phys.\ Rev.\  D {\bf 44} (1991) 587;
  P.~H.~Frampton and A.~Rasin,
  Phys.\ Lett.\  B {\bf 478} (2000) 424
[arXiv:hep-ph/9910522];
  J.~Kubo, A.~Mondragon, M.~Mondragon and E.~Rodriguez-Jauregui,
  Prog.\ Theor.\ Phys.\  {\bf 109} (2003) 795
  [Erratum-ibid.\  {\bf 114} (2005) 287]
  [arXiv:hep-ph/0302196];
  W.~Grimus and L.~Lavoura,
  Phys.\ Lett.\  B {\bf 572} (2003) 189
   [arXiv:hep-ph/0305046].

   \bibitem{ma}
  E.~Ma and G.~Rajasekaran,
  Phys.\ Rev.\  D {\bf 64} (2001) 113012
  [arXiv:hep-ph/0106291];
  K.~S.~Babu, E.~Ma and J.~W.~F.~Valle,
  Phys.\ Lett.\  B {\bf 552} (2003) 207
[arXiv:hep-ph/0206292];
  W.~Grimus, A.~S.~Joshipura, S.~Kaneko, L.~Lavoura and M.~Tanimoto,
  JHEP {\bf 0407} (2004) 078
 [arXiv:hep-ph/0407112];
  K.~S.~Babu and X.~G.~He,
  arXiv:hep-ph/0507217;
  G.~Altarelli and F.~Feruglio,
  Nucl.\ Phys.\  B {\bf 720} (2005) 64
 [arXiv:hep-ph/0504165];
  C.~Hagedorn, M.~Lindner and R.~N.~Mohapatra,
  JHEP {\bf 0606} (2006) 042
[arXiv:hep-ph/0602244];
  E.~Ma, H.~Sawanaka and M.~Tanimoto,
  Phys.\ Lett.\  B {\bf 641} (2006) 301
[arXiv:hep-ph/0606103];
  F.~Feruglio, C.~Hagedorn, Y.~Lin and L.~Merlo,
  Nucl.\ Phys.\  B {\bf 775} (2007) 120
 [arXiv:hep-ph/0702194];
  S.~F.~King and M.~Malinsky,
  Phys.\ Lett.\  B {\bf 645} (2007) 351
 [arXiv:hep-ph/0610250];
  K.~S.~Babu and S.~Gabriel,
  Phys.\ Rev.\  D {\bf 82}, 073014 (2010)
  [arXiv:1006.0203 [hep-ph]];
   For reviews see:
  H.~Ishimori, T.~Kobayashi, H.~Ohki, H.~Okada, Y.~Shimizu and M.~Tanimoto,
  Prog.\ Theor.\ Phys.\ Suppl.\  {\bf 183} (2010) 1
  [arXiv:1003.3552 [hep-th]];
   G.~Altarelli and F.~Feruglio,
  Rev.\ Mod.\ Phys.\  {\bf 82}, 2701 (2010)
  [arXiv:1002.0211 [hep-ph]].



\bibitem{FCNC}
  J.~F.~Donoghue, H.~P.~Nilles and D.~Wyler,
  Phys.\ Lett.\  B {\bf 128} (1983) 55;
  M.~J.~Duncan,
  Nucl.\ Phys.\  B {\bf 221} (1983) 285;
  F.~Gabbiani and A.~Masiero,
  Nucl.\ Phys.\  B {\bf 322} (1989) 235;
  J.~S.~Hagelin, S.~Kelley and T.~Tanaka,
  Nucl.\ Phys.\  B {\bf 415} (1994) 293;
  F.~Gabbiani, E.~Gabrielli, A.~Masiero and L.~Silvestrini,
  Nucl.\ Phys.\  B {\bf 477} (1996) 32
  [arXiv:hep-ph/9604387];
  M.~Ciuchini {\it et al.},
  JHEP {\bf 9810} (1998) 008
 [arXiv:hep-ph/9808328];
  D.~Becirevic {\it et al.},
  Nucl.\ Phys.\  B {\bf 634} (2002) 105
  [arXiv:hep-ph/0112303].

  \bibitem{dine}
  M.~Dine, R.~G.~Leigh and A.~Kagan,
  Phys.\ Rev.\  D {\bf 48} (1993) 4269
 [arXiv:hep-ph/9304299];
  R.~Barbieri, G.~R.~Dvali and L.~J.~Hall,
  Phys.\ Lett.\  B {\bf 377} (1996) 76
[arXiv:hep-ph/9512388];
  M.~C.~Chen and K.~T.~Mahanthappa,
  Phys.\ Rev.\  D {\bf 65} (2002) 053010
[arXiv:hep-ph/0106093];
  S.~F.~King and G.~G.~Ross,
  Phys.\ Lett.\  B {\bf 574} (2003) 239
[arXiv:hep-ph/0307190];
  G.~G.~Ross, L.~Velasco-Sevilla and O.~Vives,
  Nucl.\ Phys.\  B {\bf 692} (2004) 50
[arXiv:hep-ph/0401064].

\bibitem{bm}
  K.~S.~Babu and S.~M.~Barr,
  Phys.\ Lett.\  B {\bf 387} (1996) 87
[arXiv:hep-ph/9606384];
  K.~S.~Babu and R.~N.~Mohapatra,
  Phys.\ Rev.\ Lett.\  {\bf 83} (1999) 2522
[arXiv:hep-ph/9906271].

\bibitem{murayama}
Y.~Kawamura, H.~Murayama and M.~Yamaguchi,
  Phys.\ Rev.\  D {\bf 51}, 1337 (1995)
  [arXiv:hep-ph/9406245].


\bibitem{seiberg}
  P.~Pouliot and N.~Seiberg,
  Phys.\ Lett.\  B {\bf 318} (1993) 169
 [arXiv:hep-ph/9308363];
  D.~B.~Kaplan and M.~Schmaltz,
  Phys.\ Rev.\  D {\bf 49} (1994) 3741
[arXiv:hep-ph/9311281];
  L.~J.~Hall and H.~Murayama,
  Phys.\ Rev.\ Lett.\  {\bf 75} (1995) 3985
[arXiv:hep-ph/9508296];
  C.~D.~Carone, L.~J.~Hall and H.~Murayama,
  Phys.\ Rev.\  D {\bf 53} (1996) 6282
[arXiv:hep-ph/9512399];
  P.~H.~Frampton and T.~W.~Kephart,
  Int.\ J.\ Mod.\ Phys.\  A {\bf 10} (1995) 4689
[arXiv:hep-ph/9409330];
  T.~Kobayashi, J.~Kubo and H.~Terao,
  Phys.\ Lett.\  B {\bf 568} (2003) 83
  [arXiv:hep-ph/0303084];
  T.~Kobayashi, S.~Raby and R.~J.~Zhang,
  Nucl.\ Phys.\  B {\bf 704} (2005) 3
[arXiv:hep-ph/0409098];
  M.~C.~Chen and K.~T.~Mahanthappa,
  Phys.\ Lett.\  B {\bf 652} (2007) 34
 [arXiv:0705.0714 [hep-ph]];
  I.~de Medeiros Varzielas, S.~F.~King and G.~G.~Ross,
  Phys.\ Lett.\  B {\bf 648} (2007) 201
[arXiv:hep-ph/0607045];
  H.~Ishimori, T.~Kobayashi, H.~Okada, Y.~Shimizu and M.~Tanimoto,
  JHEP {\bf 0912} (2009) 054
[arXiv:0907.2006 [hep-ph]];
 L.~L.~Everett and A.~J.~Stuart,
  Phys.\ Rev.\  D {\bf 79}, 085005 (2009)
  [arXiv:0812.1057 [hep-ph]];  arXiv:1011.4928 [hep-ph].





  \bibitem{bk}
  K.~S.~Babu and J.~Kubo,
  Phys.\ Rev.\  D {\bf 71} (2005) 056006
 [arXiv:hep-ph/0411226].

 \bibitem{Lenz:2010gu}
  A.~Lenz {\it et al.},
  arXiv:1008.1593 [hep-ph].

 \bibitem{Kajiyama:2005rk}
  Y.~Kajiyama, E.~Itou and J.~Kubo,
  Nucl.\ Phys.\  B {\bf 743} (2006) 74
[arXiv:hep-ph/0511268];
  N.~Kifune, J.~Kubo and A.~Lenz,
  Phys.\ Rev.\  D {\bf 77} (2008) 076010
 [arXiv:0712.0503 [hep-ph]];
  K.~Kawashima, J.~Kubo and A.~Lenz,
  Phys.\ Lett.\  B {\bf 681} (2009) 60
 [arXiv:0907.2302 [hep-ph]].

   \bibitem{Babu:2009nn}
  K.~S.~Babu and Y.~Meng,
  Phys.\ Rev.\  D {\bf 80} (2009) 075003
[arXiv:0907.4231 [hep-ph]].


\bibitem{Araki:2008rn}
  T.~Araki and J.~Kubo,
  Int.\ J.\ Mod.\ Phys.\  A {\bf 24} (2009) 5831
[arXiv:0809.5136 [hep-ph]].



   \bibitem{UTfit}
   M.~Bona {\it et al.},
  arXiv:0906.0953 [hep-ph],
  %
  http://www.utfit.org/

   \bibitem{Kubo:2010mh}
  J.~Kubo and A.~Lenz,
  Phys.\ Rev.\  D {\bf 82} (2010) 075001
  [arXiv:1007.0680 [hep-ph]];
  Y.~Kaburaki, K.~Konya, J.~Kubo and A.~Lenz,
  arXiv:1012.2435 [hep-ph].

   \bibitem{Xing:2007fb}
  Z.~z.~Xing, H.~Zhang and S.~Zhou,
  Phys.\ Rev.\  D {\bf 77} (2008) 113016.
 [arXiv:0712.1419 [hep-ph]].

\bibitem{Nakamura:2010zzi}
  K.~Nakamura {\it et al.}  [Particle Data Group],
  J.\ Phys.\ G {\bf 37} (2010) 075021.


\bibitem{Kubo:2003iw}
  J.~Kubo, A.~Mondragon, M.~Mondragon and E.~Rodriguez-Jauregui,
  Prog.\ Theor.\ Phys.\  {\bf 109} (2003) 795
  [Erratum-ibid.\  {\bf 114} (2005) 287];
  [arXiv:hep-ph/0302196];
  J.~Kubo,
  Phys.\ Lett.\  B {\bf 578} (2004) 156
  [Erratum-ibid.\  B {\bf 619} (2005) 387]
 [arXiv:hep-ph/0309167].

  \bibitem{Schwetz:2008er}
  T.~Schwetz, M.~A.~Tortola and J.~W.~F.~Valle,
  New J.\ Phys.\  {\bf 10} (2008) 113011
  [arXiv:0808.2016 [hep-ph]].


  \bibitem{Aushev:2010bq}
  T.~Aushev {\it et al.},
  arXiv:1002.5012 [hep-ex].

  \bibitem{Ichikawa:2010zza}
  A.~K.~Ichikawa  [T2K Collaboration],
  J.\ Phys.\ Conf.\ Ser.\  {\bf 203} (2010) 012104.

\bibitem{Colangelo:2010et}
  G.~Colangelo {\it et al.},
  arXiv:1011.4408 [hep-lat].

\end{thebibliography}
\end{document}